\documentclass[conference]{IEEEtran}
\IEEEoverridecommandlockouts

\usepackage{pgfplots}

\usepackage{amsthm,amsmath,amssymb,amsfonts}
\usepackage{graphicx}
\usepackage{textcomp}
\usepackage{url}

\usepackage{flushend}
\usepackage[linesnumbered,algoruled,boxed,lined]{algorithm2e}

\newcommand{\erwan}[1]{\textcolor{red}{[\textbf{E}:#1]}}

\newcommand{\benoit}[1]{\textcolor{orange}{[\textbf{B}:#1]}}

\newcommand{\ie}{\emph{i.e.}, }
\newcommand{\eg}{\emph{e.g.}, }

\newcommand{\pol}{\texttt{DEPUTEES}\xspace}
\newcommand{\rd}{\texttt{RANDOM}\xspace}
\newcommand{\bots}{\texttt{BOTS}\xspace}
\newcommand{\fa}{\texttt{FAMOUS}\xspace}

\usepackage[makeroom]{cancel}

\usepackage{subcaption}

\theoremstyle{plain}

\usepackage{ color, colortbl}
\definecolor{Gray}{gray}{0.9}

\usepackage{tikz}
\usetikzlibrary{arrows, arrows.meta, positioning,shapes}

\usepackage{cite}
\usepackage{amsmath,amssymb,amsfonts}
\usepackage{algorithmic}
\usepackage{graphicx}
\usepackage{textcomp}
\usepackage{xcolor}
\def\BibTeX{{\rm B\kern-.05em{\sc i\kern-.025em b}\kern-.08em
    T\kern-.1667em\lower.7ex\hbox{E}\kern-.125emX}}
\begin{document}

\title{Setting the Record Straighter on Shadow Banning}

\author{
  \IEEEauthorblockN{Erwan {Le Merrer},
    \IEEEauthorblockA{Univ Rennes, Inria, CNRS, Irisa\\erwan.le-merrer@inria.fr}}
    \and
    \IEEEauthorblockN{Beno\^it Morgan,
    \IEEEauthorblockA{IRIT/ENSHEEIT\\benoit.morgan@irit.fr}}
    \and
    \IEEEauthorblockN{Gilles Tr\'edan,
    \IEEEauthorblockA{LAAS/CNRS\\gtredan@laas.fr}}
}

\maketitle

 \begin{abstract}

   \textit{Shadow banning} consists for an online social network in
   limiting the visibility of some of its users, without them being
   aware of it. Twitter declares that it does not use such a practice,
   sometimes arguing about the occurrence of ``bugs'' to justify
   restrictions on some users. This paper is the first to address the
   plausibility of shadow banning on a major online platform,
   by adopting both a statistical and a graph topological approach.

   We first conduct an extensive data collection and analysis
   campaign, gathering occurrences of visibility limitations on user
   profiles (we crawl more than $2.5$ millions of them). In such a
   black-box observation setup, we highlight the salient user profile
   features that may explain a banning practice (using machine
   learning predictors). We then pose two hypotheses for the
   phenomenon: \textit{i)} limitations are bugs, as claimed by
   Twitter, and \textit{ii)} shadow banning propagates as an epidemic
   on user-interaction ego-graphs.  We show that hypothesis
   \textit{i)} is statistically unlikely with regards to the data we
   collected. We then show some interesting correlation with
   hypothesis \textit{ii)}, suggesting that the interaction topology
   is a good indicator of the presence of groups of shadow banned
   users on the service.
  \end{abstract}

\maketitle

\section{Introduction}

Online Social Networks (OSNs) like Twitter, Facebook, Instagram or YouTube
control the visibility of the content uploaded by their users.
They have the capacity to promote or demote specific contents, and
have great responsibilities (\eg to moderate hate speech, prevent
automation for influence gain \cite{founta2018large} or to defend
copyright ownership).
OSNs often position themselves as free speech defenders.

While OSNs need to implement policies that satisfy such requirements,
precise policies are rarely publicly displayed. Therefore, debates on
their behavior with respect to some content they host is
generally fueled by three sources: \emph{i)} OSN's official statements, \emph{ii)} anecdotal evidence from users
publicizing their observations (\eg
particular requests such as "Clinton vs Trump" \cite{engadget}), and \emph{iii)}
whistle-blowing from internal sources \cite{tik} or internal information leaks.
Investigation journalism sometimes discusses the problem
in a broader context with disparate methods
\cite{diparate-journal}.

While debates about perceived freedom of speech are inevitable,
we believe it is important to explore techniques to shed light on OSNs
content regulation practices. More precisely, means to observe\footnote{We operate a shadow banning test website: \url{https://whosban.eu.org} \cite{whosban}.}, assess
and quantify the effects of content moderation is important for the
debate on information regulation in the public sphere. However, as
content is produced and consumed distributedly, and as its moderation
happens on the OSN side, collecting information about potential issues
is difficult.  In this paper, we explore scientific approaches to shed
light on Twitter's alleged \textit{shadow banning} practice. Focusing
on this OSN is crucial because of its central use as a public
communication medium, and because potential shadow banning practices
were recently commented.

\noindent\textit{Shadow banning and moderation techniques.}  shadow
banning (SB or banning for short, also known as \textit{stealth
  banning} \cite{doi:10.1177/1461444818773059}) is an online
moderation technique used to ostracise undesired user behaviors. In
modern OSNs, shadow banning would refer to a wide range of techniques
that artificially limit the visibility of targeted users or user
posts (see \eg ref. \cite{insta-sb} for a position on shadow banning
in Instagram).

Some people claim what they publish is discriminated by a moderation
algorithm \cite{doi:10.1177/1461444818773059}.  However, while
platforms publicly acknowledge the use of automatic moderation, they
deny the use of shadow banning. In particular, in a dedicated blog
post entitled \textit{``Setting the record straight on shadow
  banning''} \cite{twitterOnSB}, Twitter acknowledged some problems in
July 2018, but presented them as patched issues or bugs.

\noindent\textit{Observation in a black-box setup.}  From a
user-standpoint, observing a remote decision-making algorithm (\eg
recommending people to follow, recommending topics, or searching and
sorting users accounts), gaining some information about the OSN
moderation practices imposes a \textit{black-box interaction setup}
(see \eg refs \cite{jiang2020reasoning,influence-cookbooks,chaintreau}
for related research works).  In such a setup, the difficulty is to be
bound to observe solely input/output relations such as actions and
consequences in the OSN, and to build a relevant case from them.

We follow a statistical approach in order to \textbf{address the
  question of the plausibility of shadow banning in Twitter}. Such an
approach was also recently embraced by Jiang \& al to collect the
context of YouTube videos, in order to assess if the political leaning
of a content plays a role in the moderation decision for its
associated comments \cite{jiang2020reasoning}. The question is
addressed statistically, to validate or reject the hypothesis of bias
by YouTube.

\noindent\textit{Contributions.}  We rely on three known techniques
\cite{sb.eu} to detect Twitter users or tweets with diminished
visibility, that we implement in a full fledged scalable and automated
crawler. 
We pursue a statistical and topological perspective on shadow banning, by
comparing the plausibility of two hypotheses. More precisely, we make the following contributions:
\begin{itemize}
\item 
  We quantify the phenomenon of shadow banning on Twitter, through an
  extensive data collection and analysis campaign.
We collect the public profiles and interactions of millions
of Twitter users, as well as their shadow banning status.
\item We identify salient profile features that contribute to the
  probability to be banned, using machine learning explainable
  predictors.
\item We test the hypothesis of a random bug $H_0$: \emph{shadow
  banned users are uniformly spread among Twitter users}, which
  corresponds to Twitter's bug claim.  We show this hypothesis to be
  statistically unlikely.
\item We propose another hypothesis $H_1$: the topological
  hypothesis. It models shadow banning as an epidemic process among
  interacting users. It leverages their interaction topologies, in
  order to capture the observed effect of groups of shadow banned
  users. We show this hypothesis to better match our collected
  observations.
\end{itemize}

The remaining of this paper is organized as follows. In Section
\ref{sec:sbt}, we define how to test for shadow banning, and detail
the collection campaign we conducted to allow focusing on the shadow
banning question in Twitter. In Section \ref{sec:statistics}, we
report statistics and analyze the presence of banned profiles.  In
Section \ref{sec:features}, we have a look at which features may
predict a shadow ban status on a user. In Section \ref{sec:hypothesis}
we introduce and study our two core work hypotheses. We review Related
Work and conclude in Sections \ref{sec:related} and
\ref{sec:conclusion}. Finally, we issue a data and code availability
statement in Section \ref{sec:data}.

\section{A Data Collection Campaign for Twitter}
\label{sec:sbt}

Studying shadow banning on Twitter requires two fundamental
ingredients: first, means to detect whether a specific user profile is
banned. Second, we need to select populations on which to apply such
user-level detection. Each population should be large enough and
representative, so that conclusions drawn can be meaningful.

\subsection{Means to Assess Shadow Banning in Twitter}
\label{sec:sbtechniques}

In the context of Twitter, the notion of \textit{shadow banning} can
describe a handful of situations where the visibility of a shadow
banned user or his posts is reduced as compared to normal visibility.
The first website to provide users with the ability to check whether
they are individually shadow banned is \textit{shadowban.eu}
\cite{sb.eu}. Interestingly, its authors provided code on GitHub, as
well as explanations of techniques to assert banning facts. We
leveraged and incorporated these techniques to develop our
crawler. Here are the types of bans we consider in the paper:
 
\begin{itemize}
\item \textit{Suggestion Ban}: Users targeted by the suggestion ban are never
  suggested, as another user performs searches or mentions them in some content. This
  limits the possibility for users to accidentally reach a banned user profile.
\item \textit{Search Ban}: Users are never shown in search results, even if their
  exact user name is searched for.
\item \textit{Ghost Ban}: If a targeted user made a tweet $t$ as a new
  thread, a retweet or a reply to someone else's tweet $t'$, it is not
  shown (but is replaced by the mention "This tweet is
  unavailable"). No button allows to see it.
\end{itemize}
We declare a user to be banned if at least one of these bans holds. We
later report their precise relative occurrence in our analysis.

It is important to highlight two properties of this detection
approach. First, it does not produce false-positives (normal users
accidentally appearing as banned): detected users have actual
diminished visibility (at least at the performed crawl
time). Moreover, our detector also produces a proof allowing a human
direct confirmation of detected case. Second, these types of bans might
only constitute a subset of Twitters' banning strategy: there might be
more methods to diminish the visibility of a user and for which we do
not know any practical detector. As a consequence, the data
collection results might underestimate shadow banning on Twitter, but
not overestimate it.

\subsection{A Data Collection Campaign}
\label{sec:data}

We built a scalable crawler to retrieve user profiles, and test the
types of bans we described.  As all Twitter's users
obviously cannot be tested for banning (Twitter in Q1 2019
reported 330 millions of monthly users\footnote{\scriptsize
  \url{https://www.statista.com/statistics/282087/number-of-monthly-active-twitter-users/}}),
we resorted to the sampling of ego-graphs around selected users, which
is a common practice for studying OSNs (see \eg \cite{ego,5599247}).

In order to analyze if  banning is concerning evenly different
types of users, we selected \textit{four} types of user
\textit{populations}. We now describe these populations, and how we
extracted them from Twitter:

\paragraph{A random population}
To uniformly select a \rd population of users, we exploit a property of the Twitter API
that associates to each user a user ID randomly drawn in a finite subset of
$\mathbb{N}$. To cope with the success of that social network, this user ID space
has been resized from 32-bit to 64-bit in late 2015. 
Current user IDs seem to be still randomly drawn from this huge 64-bit space
which is for now still sparse : 330 millions over 18 billion billion, leaving us
a probability less than $1.8\times10^{-11}$ to pick an actual account at
random. Due to obvious time limitations, we decided to use the first user ID
space to draw random accounts, created before late 2015.
Therefore, our \rd population contains pre-2015 users sampled by
drawing uniformly at random  user IDs in the range $[1,2^{32} - 1]$.

\paragraph{Identified bots}

For collecting a population of \bots \cite{10.1145/3298789}, we
leveraged the website {\it \url{https://botsentinel.com}}, that has
the purpose of identifying and listing bots operating in Twitter. Bots
are classified into categories accessible using a web interface. We
have chosen to use the so called "Trollbot" category, because of their
"perceived likelihood" of being shadow banned.  We have instrumented
the HTTP REST API endpoint used by the web interface in order to
extract 1,500 account screen names.

\paragraph{Celebrities}

To build a population of very visible user accounts, we denote \fa,
we leveraged the website {\it
  \url{https://majesticmonitor.com/free-tools/social-explorer}}. This
application offers a hierarchical ranking, by topic, of the 10 most
famous Twitter accounts as follows. Topics hierarchy is a tree where
its root gathers the 10 most famous Twitter accounts from all
topics. Root siblings are in turn gathering the 10 most famous Twitter
accounts from their own topics and subtopics, and so on. That is, the
user is able to display the top ten Twitter accounts by category, by
browsing in the tree based hierarchy.

Once again we have scripted browser interactions in order to
extract the 1,500 most famous accounts, using a depth limited
breadth-first search over the HTTP REST API.

\paragraph{Political representatives in France}
We build a population we coin \pol, gathering the full list of elected
deputees who have a Twitter account in France \cite{nosdeputes} (577 as of December 2019).
We target this population because of its specific exposure to the media.

\subsection{Sampling ego-graphs in the Twitter Interaction Graph.}
Rather than simply crawling individual profiles in each of these
populations, we rely on "snowball" sampling from these
profiles to capture ego-graphs topologies around them.

More precisely, we consider the \emph{Twitter interaction graph} as
the graph $G_{Twitter}=(V,E)$ constituted by $V$ the set of all
Twitter user accounts, and $E$ a set of directed edges established as
follows: $(u,v)\in E \Leftrightarrow$ user $v$ \textit{replied} to
$u$, or \textit{retweeted} one of $u$'s messages\footnote{Note that this graph differs from the explicit Twitter
  graph in which edges capture the "follower" relationship, examined
  for instance in \cite{10.1145/2567948.2576939}.}.  As crawling the
full Twitter interaction graph is out of the question, we sample
ego-graphs from that graph as follows (see \eg ref. \cite{ego,5599247} for
other works extraction ego-graphs in the interaction graph).

We call each of the user profile in the four populations
 a \textit{landmark}, around which the ego-graph will be recursively
sampled in the interaction graph.  More
precisely, from each of these landmarks $l$, we conduct a
depth-limited breadth-first search: we parse the $33$ first tweets of
$l$ returned among its $1,000$ most recent tweets,
and list the set of users $V_{out}(l)$ with whom $l$ interacted. We
then repeat that procedure for each $i\in V_{out}(l)$, to discover the
two-hop neighbors of landmark $l, V^2_{out}(l)$. Finally, we also keep the neighbors of those rank-two nodes. The resulting ego-graph
for landmark $l$, is noted $G_l$ and is the sub-graph of $G_{Twitter}$
induced by some of its close neighboring profiles
$V_l=\bigcup_{i=1,2,3}V^i_{out}(l)$.

Note that using this process, although we chose the initial landmarks,
we do not control the population in the ego-graphs: any user
interacting with a landmark (or its ego-graph neighbors) will also
appear in its ego-graph.

Our crawling campaign took place in April 2020.
We run the set of shadow banning tests for each visited profile in
each ego-graph, on all the tweets posted since 2019 and still
available. Users that did not post any tweet since 2019 are considered
inactive and ignored for further analysis.

\subsection{Ego-Graph Collection Results.}

We targeted around $1,000$ graphs per category (except for the \pol
population that is bounded below by nature). We consider a graph to be
suitable if it contains at least two nodes (\ie the landmark and one
neighbor at the very least). As shown in the first column of Table
\ref{table:topo}, the number of graphs to extract in order to filter
out those with only one node vary greatly depending on the target
population.  Indeed, while the \pol population is very dense
(filtering graphs with one node removes only $20$ graphs over
$512$). The \rd one is crawling intensive: we needed the gather
$13,991$ graphs to be able to keep $947$ filtered ones (\ie
we filtered out the $93\%$ of users sampled at random that never interacted with someone
since 2019). 

We report no throttling nor crawl limitation from Twitter during this
data collection campaign. We performed a distributed crawling from
$86$ machines, resulting in a rate of around 100 profiles crawled per
second.
    The total amount of crawled and tested user
profiles adds up to above $2.5$ millions.

\section{Crawl Statistics: Traces of Shadow Banning}
\label{sec:statistics}

\subsection{Shadow Banning Prevalence in Populations.}

\begin{figure}[t!]
      \includegraphics[width=\linewidth]{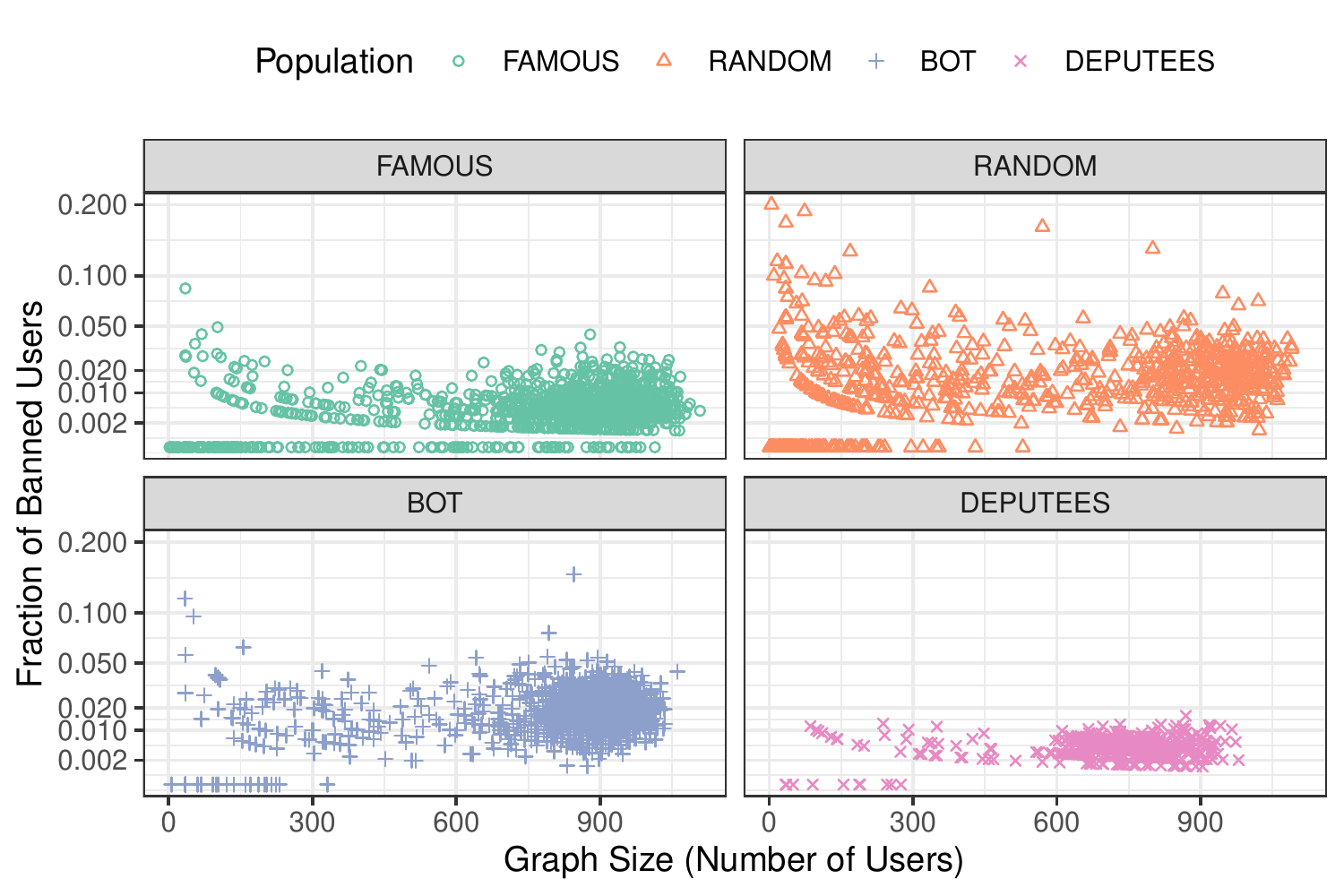}
    \caption{Fraction of shadow banned nodes ($y$-axis) as a function of ego-graph sizes ($x$-axis), for the four populations.}
    \label{fig:fractionOfBanByGraph}
\end{figure}

\begin{table*}[t!]
\begin{tabular}{|
>{\columncolor[HTML]{C0C0C0}}l |cccc}
\hline
\cellcolor[HTML]{EFEFEF} & \multicolumn{1}{l|}{\cellcolor[HTML]{C0C0C0}\#SB nodes} & \multicolumn{1}{l|}{\cellcolor[HTML]{C0C0C0}\% of SB nodes/graph (avg)} & \multicolumn{1}{l|}{\cellcolor[HTML]{C0C0C0}Degree of nodes SB|not SB (avg)} & \multicolumn{1}{l|}{\cellcolor[HTML]{C0C0C0}Fraction of SB neighbors:  node is SB|not SB } \\ \hline
\fa                   & 6,805                                                    & 0.74                                                              & 4.97 | 8.69                                                                  & 0.1044 | 0.0051                                                                      \\ \cline{1-1}
\rd                   & 9,967                                                    & 2.34                                                              & 6.94 | 9.59                                                                  & 0.1694 | 0.0211                                                                      \\ \cline{1-1}
\bots                     & 23,358                                                   & 1.97                                                              & 11.40 | 15.04                                                                & 0.0443 | 0.0184                                                                      \\ \cline{1-1}
\pol                 & 1,746                                                    & 0.50                                                              & 22.18 | 14.40                                                                & 0.0195 | 0.0104                                                                      \\ \cline{1-1}
\end{tabular}
\caption{Data collection campaign statistics: shadow banning (denoted SB) information on individual profiles, their neighbors, and their relative degree.}
\label{table:sb}
\end{table*}

We first plot in Figure \ref{fig:fractionOfBanByGraph} the fraction of
shadow banned users present in each ego-graph. We observe that all
populations are concerned by the phenomenon of shadow banning, across
all the spectrum of ego-graph sizes. The detailed statistics are
reported in Table \ref{table:sb}.  The percentage of banned users
 in populations ranges from 0.50$\%$ for \pol, to 2.34$\%$ for
\rd (and 0.74$\%$ for \fa, 1.97$\%$ for \bots). We note that the raw
number of shadow banned profiles for \fa looks relatively high (23,358),
but this is due to the high density of ego-graphs, that are making our
crawl to retrieve more than three times more profiles (1,179,949,
see Table \ref{table:topo}) than for the \rd population for instance.
Very noticeably, the \rd population is thus touched close to five
times more by the shadow banning phenomenon than the \pol
population. This already questions supposedly even spread of shadow
banning in Twitter, due to a bug for instance.

Second, we observe significantly different statistics such as the
average degrees of nodes knowing that a node is itself banned or not. In
particular, the fraction of neighbors of a banned node that are also banned is
much higher that for nodes that are not banned. This remark is consistent
across the four populations. This constitutes a first indication of
the existence of ``groups of shadow banned users'', captured by the topology of
ego-graphs.

\subsection{Co-occurrence of Types of Bans: a Graduated Response?}
We presented general statistics about shadow banned users in four different
populations. As we explained that a shadow ban status can come from three
types of bans (and at least one was sufficient so that we declare a
user as shadow banned), we now have a look at each type of ban in the dataset,
in order to question a possible Twitter shadow banning policy as a
reaction to user misbehaving.

\begin{figure}[h!]
  \centering
      \includegraphics[width=0.9\linewidth]{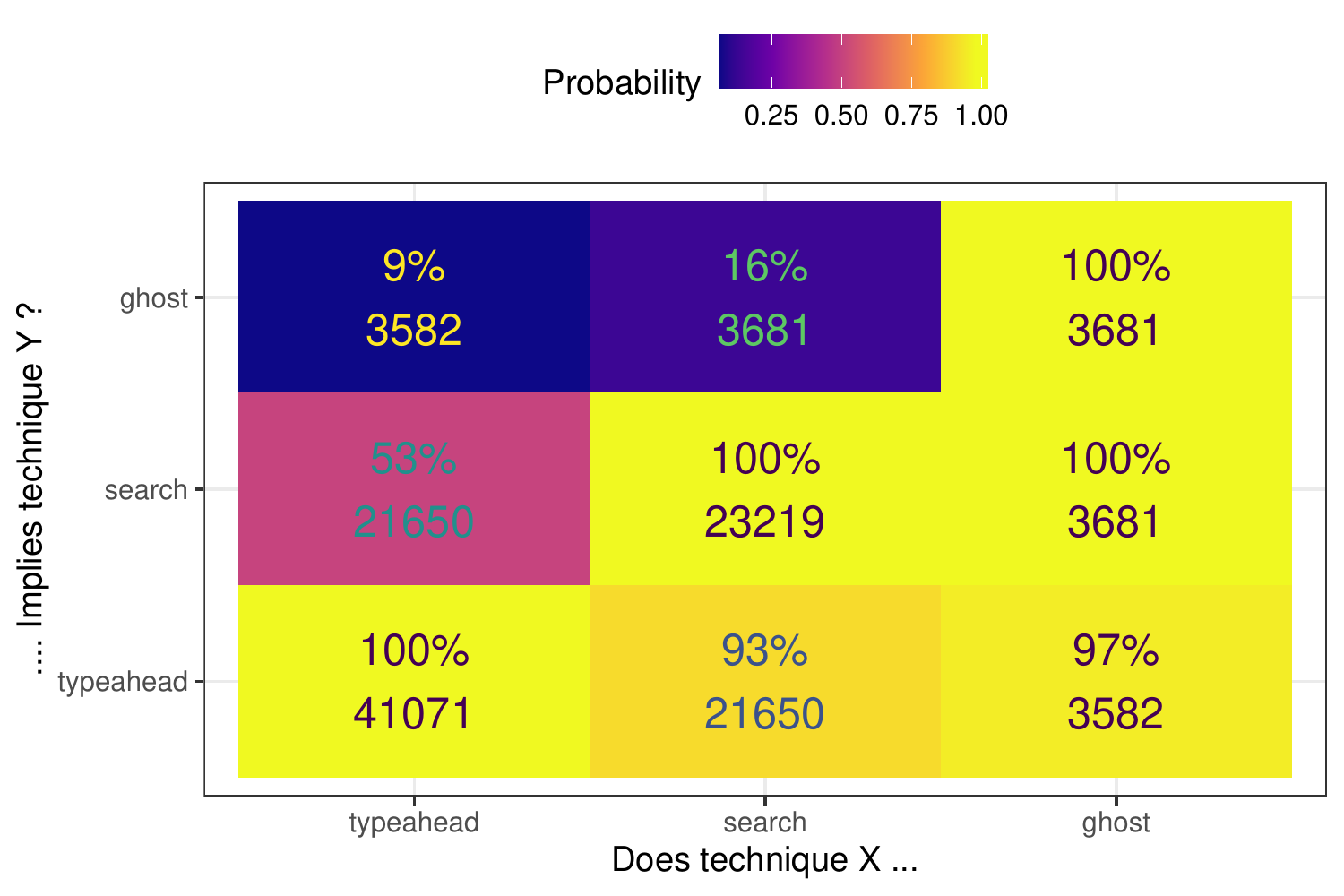}
    \caption{The interaction of the three different types of bans we
      measured in our dataset. For instance, 97\% of users that are
      ghost banned are \textbf{also} typeahead banned, while the
      reverse is true in only  9\% of the counted cases.}
    \label{fig:inter}
\end{figure}

Figure \ref{fig:inter} reports a grand total of $41,071$ \textit{typeahead} banned profiles, $23,219$ \textit{search} bans, but only $3,681$ \textit{ghost} bans.

Shadow banning techniques described in Section \ref{sec:sbt} imply
different consequences on user profiles. Their impact on visibility
can be ordered by increasing severity. As a matter of fact, {\it
  typeahead} ban is less impacting than {\it search} ban, which in
turn is less impacting than {\it ghost ban}. Indeed, the first two
sanctions leverage access to the profile while the last one the
publications themselves.

If we consider shadow banning as a punitive response against unwanted behavior as
in penal law enforcement, one could say that:
\begin{enumerate}
  \item on the one hand a punitive
reaction could be graduated according to the severity of one misbehaving;
  \item while on the other hand, recidivists could be disciplined several times,
    with increasing severity.
\end{enumerate}

Observations depicted in Figure \ref{fig:inter} seem to fit quite well
with these two points. Indeed, users solely \textit{typeahead} banned are
moderately to very little \textit{search} banned (53\%) or ghost (9\%)
banned, while \textit{ghost} banned users are almost every time search banned
(100\%) or typeahead (97\%) banned. Let us remind that our data set is
one snapshot at a given point in time. Although nicely fitting to the
collected data and the severity order, being able to observe the
evolution of sanctions per user in time would also have been of a
great interest to strongly conclude on the second point.

\subsection{Graph Topology Related Statistics.}

\begin{table*}[h!]
\center
  \begin{tabular}{|l|ccccc}
    \hline
\rowcolor[HTML]{C0C0C0} 
\cellcolor[HTML]{EFEFEF} Population        & \multicolumn{1}{l|}{\cellcolor[HTML]{C0C0C0}\#graphs | (unfiltered \#) } & \multicolumn{1}{l|}{\cellcolor[HTML]{C0C0C0}Total \#nodes} & \multicolumn{1}{l|}{\cellcolor[HTML]{C0C0C0}degree (undirected, avg)} & \multicolumn{1}{l|}{\cellcolor[HTML]{C0C0C0}clustering (avg)} & \multicolumn{1}{l|}{\cellcolor[HTML]{C0C0C0}2-core size (avg)} \\ \hline
\cellcolor[HTML]{C0C0C0}\fa   & 1211 | 1400                                                  & 908,131                                                          & 5.92                                                                  & 0.2074                                                        & 575.64                                                         \\ \cline{1-1}
\cellcolor[HTML]{C0C0C0}\rd   & 947 | 13,991                                                   & 424,489                                                          & 5.45                                                                  & 0.1965                                                        & 357.21                                                         \\ \cline{1-1}
\cellcolor[HTML]{C0C0C0}\bots     & 1436 | 1505                                                 & 1,179,949                                                        & 11.62                                                                & 0.1754                                                        & 751.64                                                         \\ \cline{1-1}
\cellcolor[HTML]{C0C0C0}\pol & 492 | 512                                                  & 348,640                                                          & 12.21                                                                 & 0.1694                                                        & 658.99                                                         \\ \cline{1-1}
\end{tabular}
  \caption{Statistics for the crawling campaign: topological
    information for extracted ego-graphs.}
  \label{table:topo}
\end{table*}

\begin{figure}[h!]
    \includegraphics[width=\linewidth]{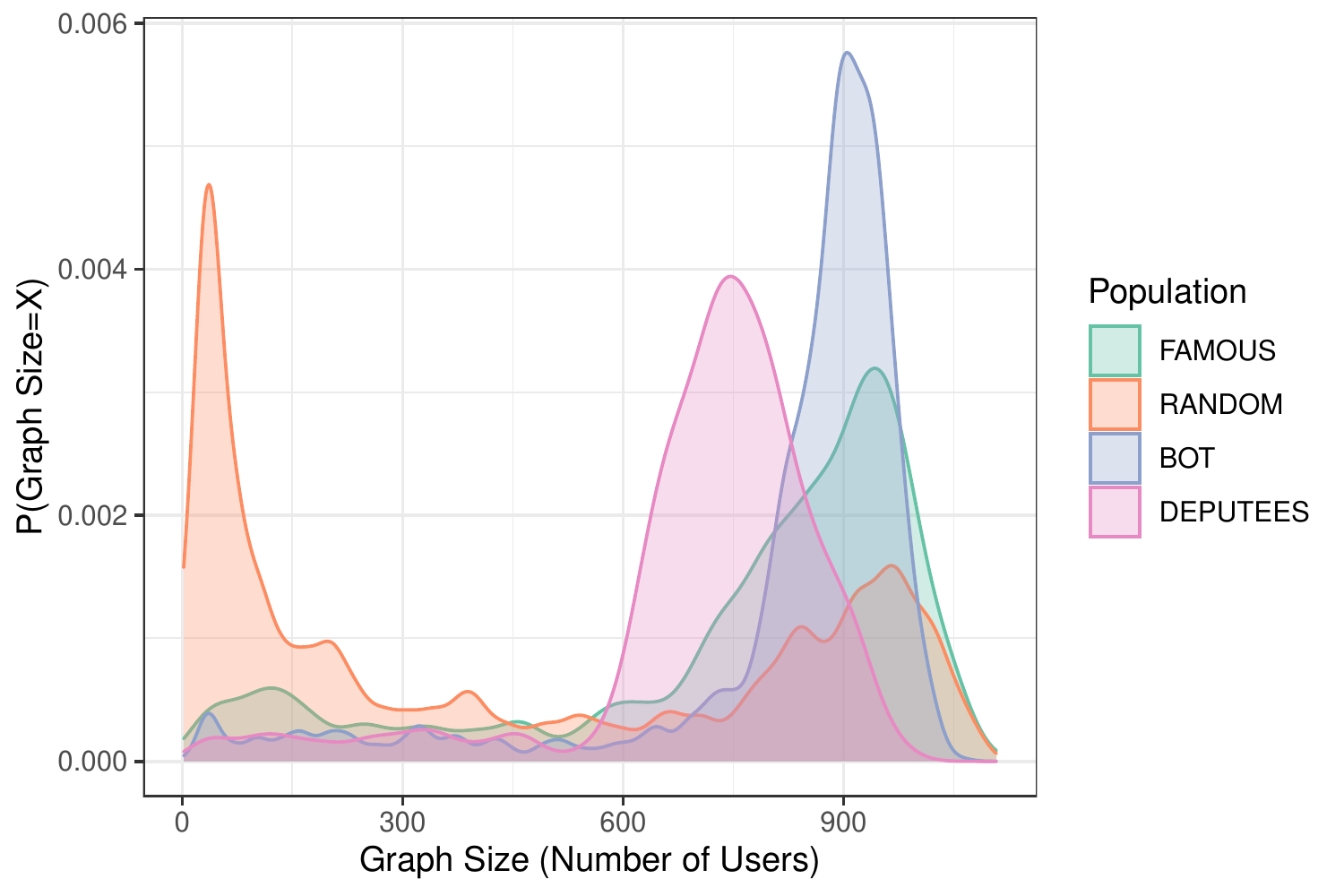}
\vspace{-.7cm}  \caption{A probability density function for the
  size of collected ego-graphs, for each of the four
  populations.}
\label{fig:sizes}
\end{figure}

We now present general statistics on the ego-graphs.
\paragraph{Graph sizes} Due to our ego-graph sampling strategy, the
size of the graphs we extract is upper bounded to
$1 + 33 + 33^2 = 1123$ nodes (corresponding to a depth two crawl
around the landmark, with $33$ neighbors at maximum per node). We
observe in Figure \ref{fig:sizes} the sizes of the collected graphs,
per population, as a probability density function. \bots and
\pol exhibit a single mode, while \fa and most notably \rd have two
modes, consisting in a fraction of graphs with small sizes and another
one of close to maximal sizes (centered at around $1,000$ nodes).

\paragraph{General graph statistics}

General statistics are reported in Table \ref{table:topo}; note that
we made all graphs undirected in our analysis, allowing for some
topological computations and analysis such as $k$-cores.  Because of
our ego-graph sampling strategy, the average degree of nodes is
expected in $[2,34]$.  Indeed, if the rank-2 nodes have only neighbors
outside the set of nodes already crawled, we have $33 + (1+33)\times33
+ 1\times33^2 = 2244$ edges at maximum; this leads to $2244/1123
\approx 2$ as an average graph degree. At the other extreme, since we capture only $33$ interactions at maximum, maximum degree is 34.
We observe in \ref{table:topo} that those averages lie
close to the lower bound for the \fa and \rd populations (with 5.92
and 5.45 respectively), and a significantly above for the \bots and
\pol populations (11.62 and 12.21).  We note the average clustering
coefficients to be relatively even (0.2074 for the largest one in the
\fa case). These statistics indicate various levels of interactions in
the four different populations; we shall explore the effect of these
interactions on the banning process.

\section{User-Features Correlating with Banning}
\label{sec:features}

We are now questioning if some features in the collected individual
user profiles are good predictors of a potential shadow ban status.

We leverage machine learning classifiers: the idea being that if one
can predict with some reasonable accuracy if a profile is shadow
banned by only looking at its features, then these features are
encoding a part of the cause of a profile being banned.  We choose three
machine learning models that are explainable \cite{molnar2019} by
construction, that is to say that the model allows for precisely
pinpoint the influence of features on the classification accuracy.
Here is the considered setup.

\paragraph{Prediction setup}
In order to train a predictor for shadow banning, we first need a labeled
dataset. A first difficulty is however the unbalanced nature of the
classification task at hand: over $97\%$ of our dataset are negative
instances (representing users that are not banned). Thus, a trivial
classifier predicting "not banned" for any input would have a $97\%$
accuracy, without bringing any information on relevant features. To
circumvent this, we first balance the dataset.

\begin{figure}[h!]
  \centering
      \includegraphics[width=1.05\linewidth]{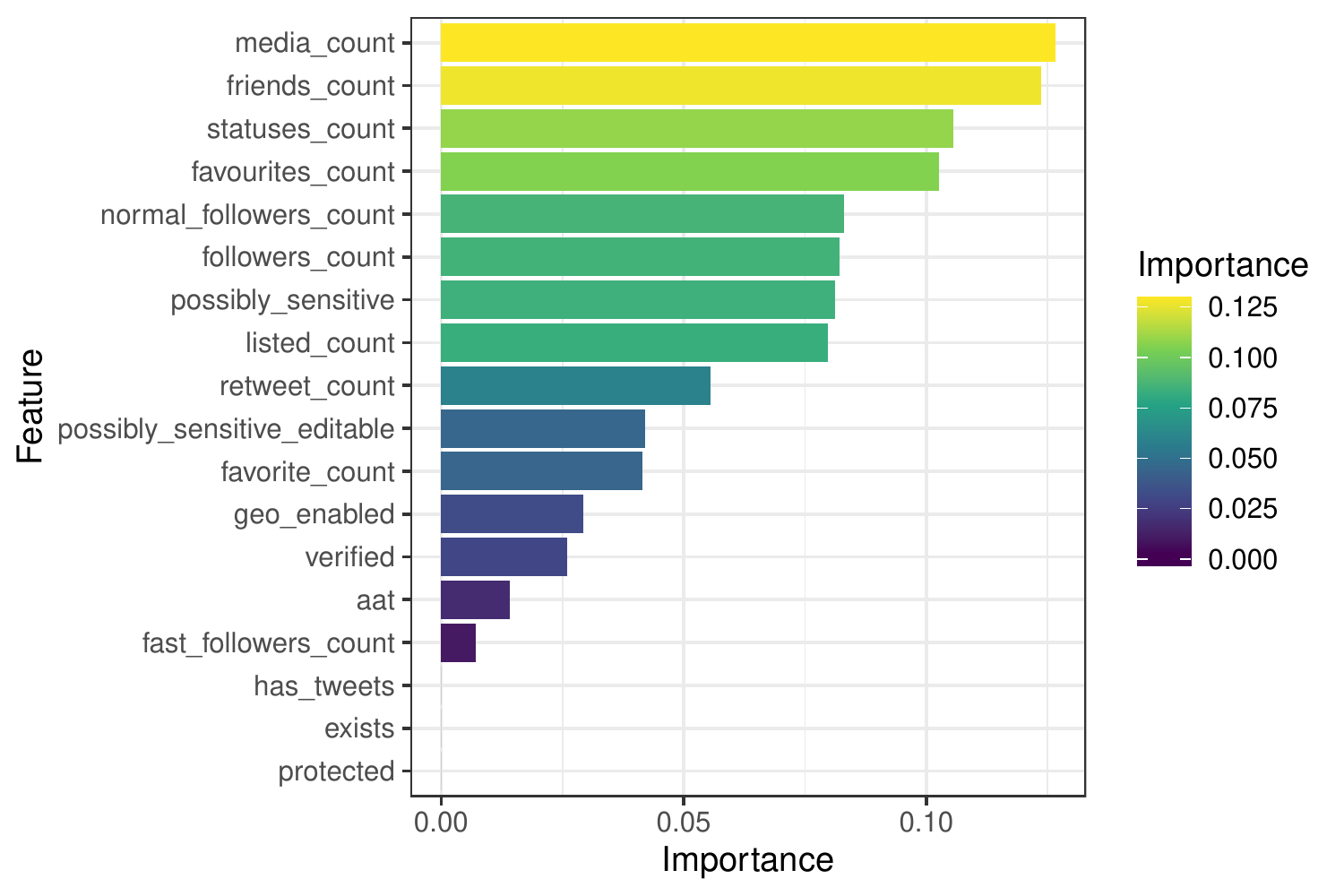}
    \caption{The features sorted by order of importance in the random
      forest model (RF) prediction of shadow banned users.}
    \label{fig:featuresML}
\end{figure}

We retain a total of $9,626$ profiles of the \rd population, all
having all of the features we leverage (we note that a large set of
profiles have unset or missing features). The shadow banned and non
shadow banned profiles are in an even quantity in the resulting
dataset; these two sets constitute our labeled dataset.
For each profile in these two sets, we use as features the data
extracted from each user Twitter webpage that is either of a Boolean
or integer format. In total, we exploit a set of 18 features that are
listed on Figure \ref{fig:featuresML} and analyzed hereafter.  The
naming of some of those data fields is very explicit (such as
\texttt{followers\_count} for instance), while some others are not
(\eg \texttt{possibly\_sensitive\_editable}).

We use the Scikit learn \cite{scikit-learn} library, and experiment with three
explainable classifier models: a random forest algorithm (RF), the
AdaBoost algorithm (AB), and a decision tree (DT). The RF is the
result of a grid search on the best combination of the following
parameters: 'number of estimators' $\in [50, 150, 250]$, 'max
features' $\in [sqrt, 0.25, 0.5, 0.75, 1.0]$ and 'min samples split'
$\in [2, 4, 6]$, leading to a optimum setup of respectively 150,
$sqrt$ and 2.  The AB is the result of a grid search on the best
combination of the following parameters: 'number of estimators' $\in
[50, 150, 250, 500]$, 'learning rate' $\in [0.1, 1, 2]$, leading to a
optimum setup of respectively 500 and 1.  Finally, the DT is the
result of a grid search on the best combination of the following
parameters: 'max features' $\in [auto, sqrt, log2]$, 'min
samples split' $\in [2\ldots 15]$, and 'min
samples leaf' $\in [1\ldots 11]$, leading to the
selection of respectively $log2$, 13 and 11.

The training is set to $80\%$ of the dataset, leaving $20\%$ of
profiles as a test set.

\paragraph{Predictor accuracies} Accuracies of the three models are reported in Table
\ref{table:accuracies}.  An accuracy of 80.6\% is observed for the RF
model (76.6\% for AB and 74.8\% for DT), which clearly shows that there
is some information in the features we collected that correlate with
the shadow banned status of tested profiles. We believe this raw result to be
encouraging for later research on even higher accuracies, for allowing
for instance services like \textit{shadowban.eu} \cite{sb.eu} or \textit{whosban.eu.org} \cite{whosban} to rely on direct
inference on public profile data, rather than on the interaction with
the Twitter services to test the ban types described in Section
\ref{sec:sbtechniques}.

\begin{table}[t!]
\center
  \begin{tabular}{|l|c}
    \hline
\rowcolor[HTML]{C0C0C0} 
\cellcolor[HTML]{EFEFEF} Classifier        & \multicolumn{1}{l|}{\cellcolor[HTML]{C0C0C0}Banned Status Prediction Accuracy} \\ \hline
\cellcolor[HTML]{C0C0C0}Random Forest (RF)   &  0.806 \\ \cline{1-1}
\cellcolor[HTML]{C0C0C0}AdaBoost (AB)   & 0.766 \\ \cline{1-1}
\cellcolor[HTML]{C0C0C0}DecisionTree (DT)   & 0.748 \\ \cline{1-1}
\end{tabular}
  \caption{Accuracies of three explainable classifiers predicting the
    shadow banned status of $1,925$ test users, based on their crawled
    profiles.}
  \label{table:accuracies}
\end{table}

\paragraph{Most salient features}
We now look at the features that are influencing the classifications
of the RF model, as it has the best accuracy and is explainable. The
relative contribution of each individual feature to the RF model
decision is represented in Figure \ref{fig:featuresML}.

We first note that there is no single feature that can help
differentiate between banned and non banned users: the decision might
be a complex combination of several of them.

There are two features with above each $12\%$ of influence on the
result: \texttt{media\_count} and \texttt{friends\_count}; above 10\%
are also two more: \texttt{statuses\_count} and
\texttt{favorite\_count}. Together those 4 features determine nearly
half of the decision ($45.8\%$). These features relate to a sort of
acceptance from general users of the user under scrutiny. This
acceptance could lead to a good indicator on the probability to be
shadow banned.

We note that the second classifier in accuracy, AB, ranks four
features over $10\%$ of influence. By decreasing order:
\texttt{media\_count}, \texttt{listed\_count},
\texttt{normal\_followers\_count} and \texttt{statuses\_count}. While
the first feature is in a rank agreement with RF, the others are not,
possibly indicating some redundancy of information in these features.


We thus conclude that despite there is no obvious public group of
features that permit perfect inference on the shadow ban status of
users, a prediction of 80\% indicates the presence of relevant information
in our crawl. (We make this crawl public for further research.)

\section{Two hypothesis: bugs and topological effect}
\label{sec:hypothesis}

The previous sections exploited the collected data at the individual
user level. Interestingly, it also revealed that at the global scale,
different populations are differently impacted by shadow banning. In
other words, banning does not appear as homogeneous, but rather
concentrated in some regions of the interaction graph.
We next seek to confirm this intuition.

\subsection{Hypothesis $H_0$: the Plausibility of Bugs}
\label{sec:indiv}

We recall hypothesis \textbf{$H_0$}: \emph{shadow banned nodes are
  uniformly distributed among Twitter users.} In this hypothesis, each
user is banned with a uniform probability $\mu$, the only
parameter of this model $H_0(\mu): \mathbb{P}(x\in SB)=\mu$.

To avoid sampling biases, we now focus on the $400,000+$ profiles \rd
population in the remaining of this paper (as the three other
populations pertain to targeted sampling of specific populations).

Fitting $H_0$ is trivial: $H_0(\mu)$ is most likely given our
observations when $\mu$ is set to be the fraction of observed banned nodes in
\rd (aka sample mean) that we write $\hat{\mu}=0.0234$ (see Table
\ref{table:topo}).
When the context is clear, we write
$H_0$ as a shorthand notation for $H_0(\hat{\mu})$.  This hypothesis embodies
the \emph{bug} explanation: bugs (software faults) are often considered to
randomly affect users \cite{faulttree}.

As $H_0$ completely ignores the topological dimension of collected
ego-graphs, what remains is a balls and bins sampling process: we can
assess the probability of observing the amount of shadow banned nodes
in each graph we collected, under $H_0$. In a nutshell in this
hypothesis, Twitter is a big bin containing a fraction $\hat{\mu}$ of
banned balls, the rest being non-banned balls. In this $H_0$
perspective, every time we sample a landmark $l$ and its ego-graph
$G_l$, we draw $|G_l|$ balls from the bin and count how many banned
balls we have drawn. In other words, every ego-graph $G_l$ we sample
is considered as $|G_l|$ realizations of Bernoulli process of
probability $\hat{\mu}$.

\begin{figure}
    \includegraphics[width=1.05\linewidth]{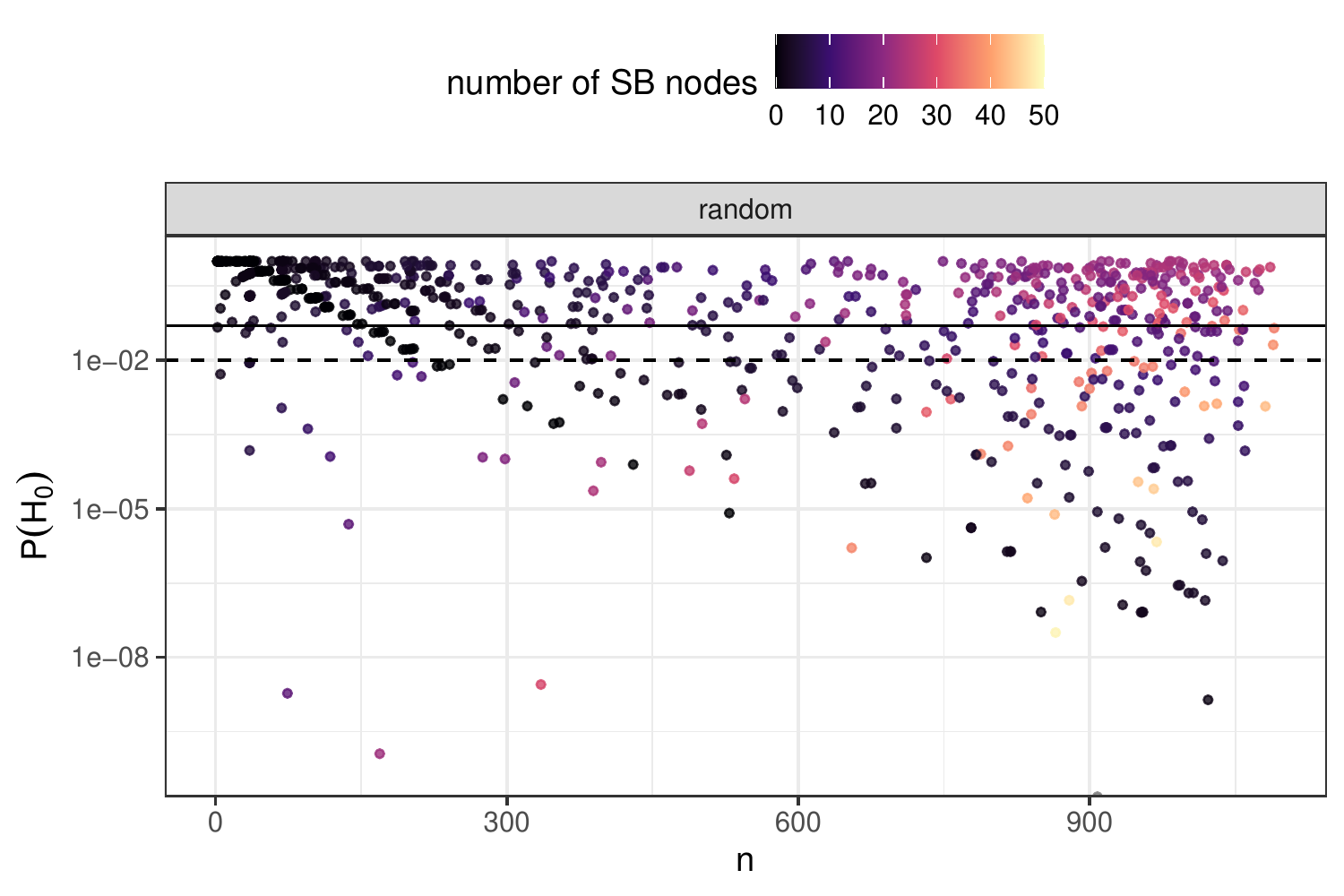}
\vspace{-.7cm}  \caption{The $p$-value of the $H_0$ (bug) hypothesis for each
    landmark. Dashed and continuous lines represent the
    $1\%$ and $5\%$ significance levels, respectively. }
    \label{fig:pvalues}
\end{figure}

We borrow a general statistical significance testing approach, as \eg
used also in ref. \cite{insta-sb}.
Given $\hat{\mu}$, estimating the probability to observe $|SB|$
successes (in other words, it's associated $p$-value) is a
 process known as Binomial test. 

Figure \ref{fig:pvalues} displays the corresponding $p$-value of
$H_0$, with regards to the size of each ego-graph and the number of
shadow banned nodes it contains. In other words, it represents the
probability that a Bernoulli trial with a success probability of
$\hat{\mu}$ leads to the number of banned user observed in each
ego-graph.  Remember that the lower the $p$-value, the higher the
plausible rejection of the hypothesis under scrutiny. We observe an
important amount of graphs that are significantly below significance
levels of $1\%$ and $5\%$: those are unlikely events to be observed
under $H_0$. Moreover the number of banned nodes in each ego-graph
(represented by point color) hints two types of unlikely ego-graphs:
large graphs with too few banned nodes (black dots) and graphs with
too many banned (clearer dots): there exist important groups of banned
nodes in some of the crawled ego-graphs.

The $y$-axis in Figure \ref{fig:pvalues} is cut under the probability
1e-09, for readability. We omitted $14$ samples that are even below
this probability. Table \ref{table:topbans} represents the top-5 most
unlikely ego-graphs we observed, around user profiles collected in our
crawl.  User account names are truncated for privacy reasons. We
observe that the ego-graph (of size 703) of user Artem* contains
45.4\% of shadow banned nodes; the likelihood of such a realization
under the $H_0$ model is 1.26e-315. In other words,
  observational data does not support hypothesis
  $H_0$.   This conclusion calls for alternative models, such as
$H_1$, which we now introduce.

\begin{table}[t!]
\center
  \begin{tabular}{|l|ccc}
    \hline
\rowcolor[HTML]{C0C0C0} 
\cellcolor[HTML]{EFEFEF}         & \multicolumn{1}{l|}{\cellcolor[HTML]{C0C0C0}Ego-graph size} & \multicolumn{1}{l|}{\cellcolor[HTML]{C0C0C0}Ratio of SB nodes} & \multicolumn{1}{l|}{\cellcolor[HTML]{C0C0C0}Probability under $H_0$}  \\ \hline
\cellcolor[HTML]{C0C0C0}Artem*   & 703                                                  & 0.454                                                          & 1.26e-315                                                                                                               \\ \cline{1-1}
\cellcolor[HTML]{C0C0C0}Vlman*   & 605                                                   & 0.443                                                          & 6.42e-262                                                                                                             \\ \cline{1-1}
\cellcolor[HTML]{C0C0C0}santi*     & 937                                                 & 0.331                                                        & 1.67e-255                                                                                                           \\ \cline{1-1}
\cellcolor[HTML]{C0C0C0}Brows* & 796                                                  & 0.241                                                          & 3.03e-130                                                                                                                 \\ \cline{1-1}
\cellcolor[HTML]{C0C0C0}ZchBr* & 763                                                  & 0.227                                                          & 9.97e-113                                                                                                                 \\ \cline{1-1}
\end{tabular}
  \caption{Top-5 most unlikely collected ego-graphs under hypothesis $H_0$, in the \rd population (along with their precise probability of observation).}
  \label{table:topbans}
\end{table}

\subsection{Hypothesis $H_1$:Interaction Topology Reflects Banning}
\label{sec:statanal}

We have concluded that our shadow banning observations do not support
the hypothesis of a random bug. Indeed, instead of revealing isolated
cases evenly scattered in different landmarks, reveals that banned
users are more concentrated around some landmarks and rarely around
some others. As the ego-graphs observed around landmarks are in
specific regions of the Twitter interaction graph, one can suspect a
relation between the topology of the interaction graph, and the
prevalence of banned users.

To investigate this, we propose an alternative hypothesis, $H_1$, that
seeks to measure how local (with respect to the interaction topology)
is the banning phenomenon. We first fit this probability, and then
inspect its likelihood w.r.t. $H_0$.

\paragraph{A simple Susceptible/Infected epidemic model}
We propose to adapt a simple Susceptible/Infected (SI) epidemic model
\cite{SI}.  Epidemic models, aside their obvious relevance in
infectiology, are widely used to describe different topologically
related phenomenons in social networks, such as information cascades
or rumor spreading \cite{10.1145/2501025.2501027}. While shadow
banning is arguably a different phenomenon that the fact of being contaminated
by a rumor, we believe the SI model to be the simplest way to capture
the intuition that some groups of interacting users are differently
touched by shadow banning.

The simplest SI model is a one step contamination process: each node is
initially infected with probability $p_0$; then, initially infected
nodes can contaminate each of their neighbors with probability
$\beta$. Therefore, this contamination process $SI$ has two parameters:
$\beta$, that captures the locality of the phenomenon, and $p_0$ that
allows to initially and uniformly spread the shadow ban status.

Let $SI(p_0,\beta)$ be our contamination process.  First, observe that
$SI(p_0=\mu,\beta=0)=H_0(\mu)$: neutralizing contamination yields the
random uniform spread of banned nodes described in $H_0$. As $\beta$ increases,
local contaminations occur around each initially infected user, and
the overall number of banned users increases.

\begin{figure}
  \begin{subfigure}[b]{0.5\textwidth}
    \includegraphics[width=\linewidth]{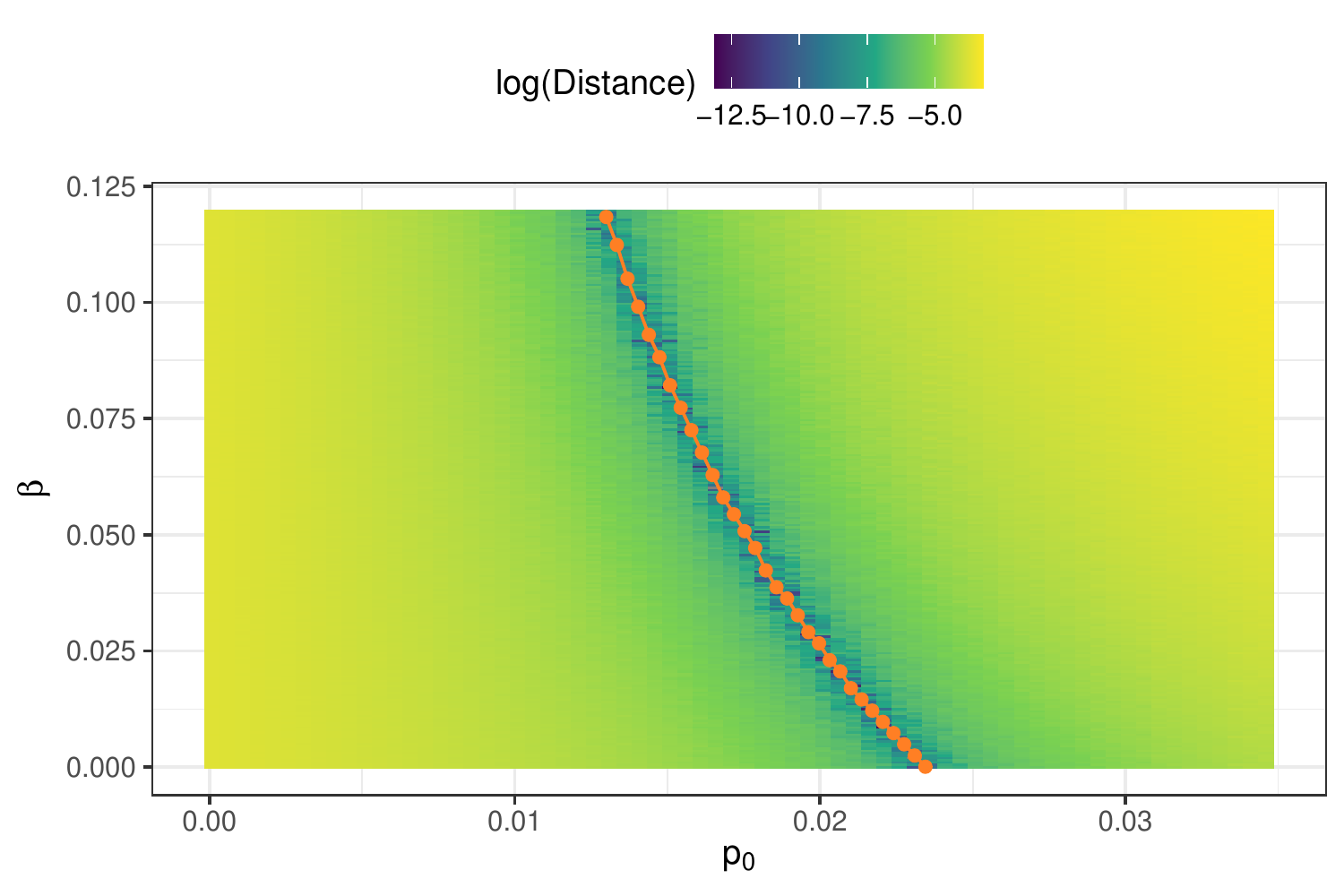}
\vspace{-.6cm}       \caption{}
    \label{fig:ridge}
  \end{subfigure}
  \begin{subfigure}[b]{0.5\textwidth}
    \includegraphics[width=\linewidth]{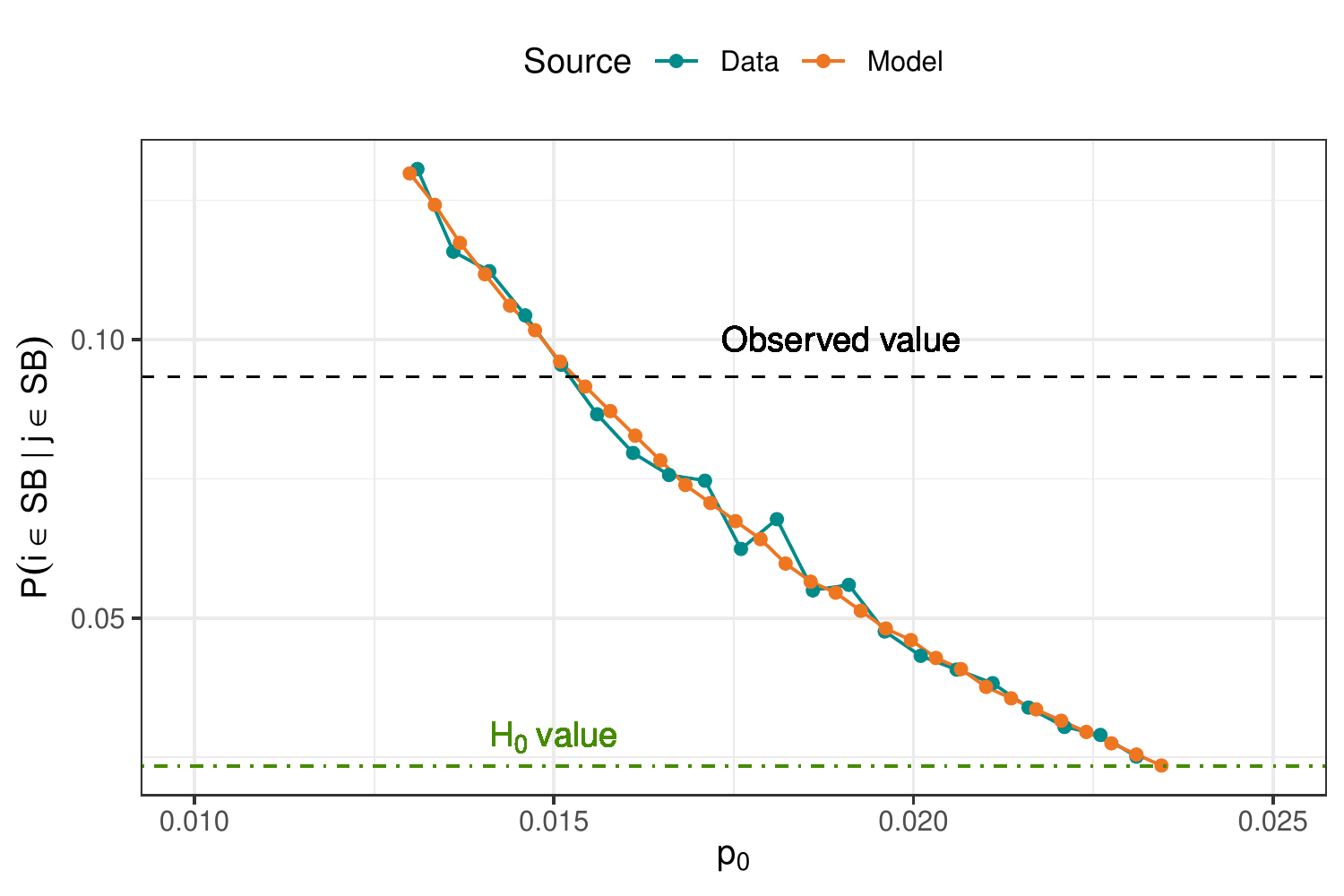}
\vspace{-.6cm}       \caption{}
    \label{fig:model}
  \end{subfigure}
\vspace{-.7cm}     \caption{  (a) The impact of $p_0$ and $\beta$ SI
  parameters on the distance of simulated shadow banned user w.r.t. to the actual
  shadow banned users in ego-graphs. The line in green corresponds to a simple
  analytical model we propose. (b) Probability of neighboring
  contamination as a function of $p_0$ for the $H_1(\beta)$ model family.}
\end{figure}

\paragraph{Fitting $H_1$ to the observations}

We seek a couple $(p_0,\beta)$ of parameters for $SI$ that are the
most likely given by our observations. A first observation is that such
likely parameter couple should reproduce the global fraction of
observed banned users $\hat{\mu}$. As observed above, $(\hat{\mu},0)$
is one such couple, but by balancing differently initial infection and
contaminations, it is possible to generate an infinity of such
couples.  Let $H_1(\beta)=SI(p_0,\beta) $ such that
$ \mathbb{P}(S| H_1(\beta))=\hat{\mu}$.

Let $S$ be the random variable associated to the event "the user is banned".
An estimation of the relation between $\mu,\beta$ and $p_0$ can be
sketched as follows: $\mathbb{P}(S| H_1(\beta))\approx
\mathbb{P}($infected initially$)\oplus
\mathbb{P}($contaminated$)=p_0+(1-p_0)p_1$. Where $p_1$ is
approximated as the probability of having some infected neighbors in a
regular random graph of degree $k$ and being contaminated by at least
one of these: $$p_1=\sum_{v=1}^k \binom{k}{v}
p_0^v(1-p_0)^{k-v}(1-(1-\beta)^v).$$ In other words, this estimation
neutralizes topological artifacts like clustering or degree
heterogeneity to sketch a rough relation between $p_0$ and $\beta$ for
a fixed $\mu$.

Figure \ref{fig:ridge} represents the quantity
$|\mathbb{P}(SB|SI(p_0,\beta)) -\mu|$ for varying $p_0$ and
$\beta$. More precisely, each point $(p_0,\beta)$ on the figure
corresponds to a SI model. We simulate this corresponding SI model on
each crawled ego-graph, and count the simulated number of banned
users. The color of the point corresponds to the difference between
the simulated number of banned users and the (real) observed number of
banned users, averaged over all ego-graphs.  In other words that is
the difference between fractions of shadow banned nodes observed on
the ego-graphs and simulated using the SI model. A distance of 0 thus
indicates that the SI simulation over the ego-graphs leads to the same
amount of banned nodes that the one counted in the dataset. A
darker color indicates a smaller distance.

We observe a smooth ridge, linking $\beta$ and $p_0$.  Note that, as expected, a
$\beta=0$ leads to $p_0$ being equal to the measured $\hat{\mu}$, that is the
initial probability to be infected, without any contamination from
neighbors.

The orange line represents the values derived from our analytical
approximation. It follows closely the lowest experimental values that
shape a valley, indicating that our model captures the process
very well.
As a consequence, the lowest spots of the valley and the orange line
both define here a family of hypotheses $H_1(\beta)$ in which all
members approximate the total number of banned nodes as closely as the
uniform infection $H_0=H_1(\beta=0)$. A natural follow-up question is
"What would be a good value for $\beta$?".

\paragraph{Probability of a banned neighbor given a banned status}
Recall that $\beta$ is the contamination probability, which is a local
property. To estimate a good value, one can look at the probability
that a banned node has a banned neighbor:
$\mathbb{P}( j\in SB |i\in SB \wedge (i,j)\in E )$. While in $H_0$
this probability is $\hat{\mu}$ (as both events $i\in SB$ and
$j\in SB$ are independent), in $H_1(\beta>0)$ the contamination
drastically increases this probability. It can be roughly estimated as
$\mathbb{P}( j\in SB |i\in SB \wedge H_1(\beta))\approx
p_0+(1-p_0)\beta$ by again neglecting clustering in ego-graphs (and chances that two
nodes contaminated by the same node are neighbors).

The empirical value we measured in the dataset for that probability is
$9.3\%$: A user having at least one banned neighbor has nearly $4$
times more chances of being banned. This observation again weakens a scenario such as
$H_0$.

Figure \ref{fig:model} represents the probability of being banned if
one has a banned neighbor for the family of $H_1(\beta)$ hypotheses
obtained above. We note a very good fit of the SI model
with the measurements from the dataset.  As expected, as $p_0$
decreases, $\beta$ increases, which in turn increases neighboring
contamination chances. The dashed line represents the empirical observed value
$|(SB\times SB) \cap E|/|(SB\times V)\cap E|$.

The model closest to this experimental line is
$H_1(\hat{\beta}=0.0955)$ corresponding to the SI model where $p_0$ is
just above $0.015$. This model would explain both the global number of
shadow banned nodes, and the local co-occurrences of shadow banning in
the data. In the following, we set $H_1$ to represent our fitted
values: $H_1:=H_1(\hat{\beta})=SI(0.015,0.0955)$.  In this model,
contaminations are $(\hat\beta)/0.015=5.4$ times more likely to occur
through neighbor contamination than through initial (random)
contamination.  We now can evaluate the likelihood of this new
hypothesis.

\subsection{Comparing the Likelihood of Observations in $H_0$ and $H_1$}

In order to conclude on both hypotheses, and to compare the occurrence
of observations in both of them, we must estimate the likelihood of
$H_1$.  Thanks to its simplicity, assessing the likelihood of $H_0$
given our observations was simple; it is not the case for $H_1$, as
one has to handle the exact impact of the topology on neighbor
contaminations. 

To circumvent this difficulty, we again resort to numerical
simulations.  On each of the $9,967$ ego-graph topologies, we simulate
$10,000$ $H_1$ model infections to estimate the resulting number of
contaminated nodes. More precisely, for each ego-graph $G_l$, let
$S_l$ be the random variable representing the number of banned nodes
obtained by simulating $H_1$ of $G_l$, and let $\hat{s_l}=|\{i\in
SB,\forall i \in V(G_l)\}|$ the observed number of banned users in
$G_l$. By simulation, we experimentally sample the probability
density function of $S_l$ and retain the probability of having exactly
$\hat{s_l}$ observations: $P(S_l=\hat{s_l}|H1)=\mathcal{L}(H_1|G_l)$,
which is likelihood of model $H_1$ on $G_l$.

\begin{figure}
    \includegraphics[width=1.03\linewidth]{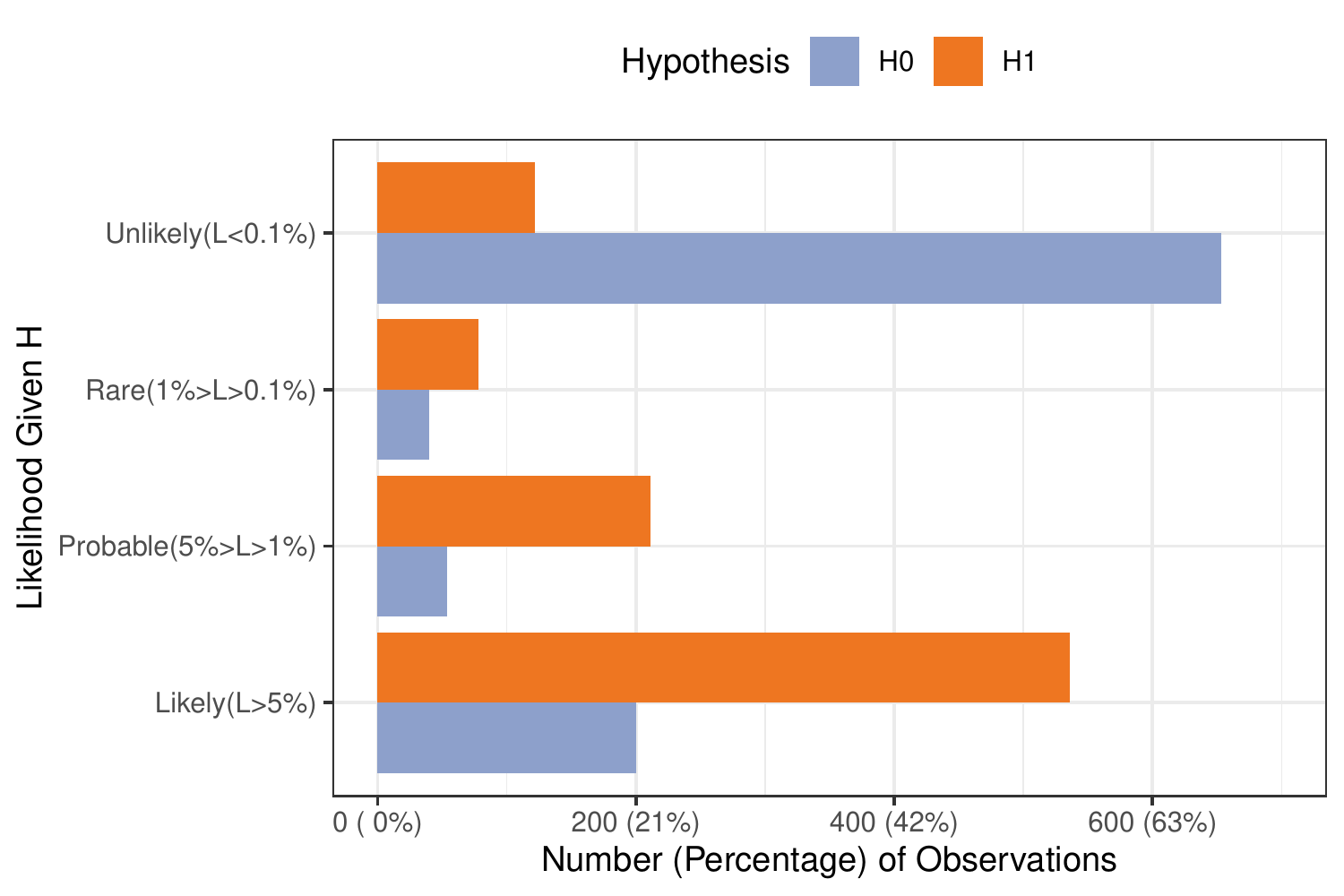}
\vspace{-.3cm} \caption{Likelihood for observing the shadow ban
  statuses in the \rd population, binned by plausibility, under both the $H_0$ and $H_1$
  hypotheses. The $H_1$ hypothesis accounts for much more likely events than the $H_0$ hypothesis, and much less unlikely events. This underlines that the topology of the interaction ego-graphs correlates with in the shadow banning phenomenon.}
    \label{fig:H1vsH0}
  \end{figure}

Figure \ref{fig:H1vsH0} reports these results. To compare the
likelihood of $H_1$ and $H_0$, we bin the likelihoods into classes of
probability occurrences for both hypotheses. This allows for a
  fair comparison of $H_1$ with $H_0$: because we resort to numerical
  evaluation of $\mathcal{L}(H_1|G_l)$, we cannot estimate by sampling
  the very low likelihoods (\eg $L<1e-4$).

Results show that likely observations under $H_1$ occur 2.68 times more.
Conversely, unlikely observations occur 5.35 times more in $H_0$
than in $H_1$.  This stresses that the $H_1$ hypothesis manages to
capture a part of what is at stake in the shadow banning process in
Twitter: the topology of ego-graphs, that is the interactions of
users, is at play.

To conclude, we have seen that hypothesis $H_0$ is unlikely. We propose
an alternative hypothesis $H_1$, that captures the locality of ban
observations with respect to the interaction graph. This model is
substantially more likely than $H_0$, revealing the tight
relation of shadow banning with respect to locality. In other words, bans appear
as clusters in certain areas typically referred to as communities in
the context of Twitter.

\vspace{0.2cm}
\section{Related Work}
\label{sec:related}
 
\paragraph{Moderation in online social networks}
Moderation of user contributions dates back to early Internet forums
such as USENET \cite{usenet}: moderation was then defined with a
parallel to ``professional journal editors'' on users
contributions. This practice is now widely spread in modern platforms
under many different shades: through a survey of 519 users who have
experienced content moderation,
West. Authors in \cite{doi:10.1177/1461444818773059} explore users' folk theories
of how content moderation systems work.  Another user-sided approach
to try to infer properties of moderation in a major OSN is conducted
in a recent short paper \cite{jiang2020reasoning}; authors are asking
the question of a potential political bias in the moderation of
YouTube comment, and forms two null hypothesis to examine it. They
leverage message content; their conclusion is that bias is supported
by one hypothesis, but not by the second.  Looking at moderation under
another aspect, Dosono et al. \cite{10.1145/3290605.3300372} study how
moderators shape communities on Reddit.

The question of automatic moderation is of interest for operators to
sustain the mass of user-produced information now available
\cite{10.1145/2189736.2189741} (here leveraging the semantics of usage
and content).

Our work differs from studies on moderation, since Twitter denies the
use of shadow banning as described in Section \ref{sec:sbt}, so we
had to develop hypotheses to check their plausibility.

\paragraph{Black-box observation of a remote service}
There is a growing literature interested in the means to extract or
infer properties from a remotely executing algorithm, from a user
standpoint. In the context of online ads, the XRay~\cite{chaintreau}
approach proposes a Bayesian angle for inferring which data of a user
profile, given as an input, is associated to a personalized ad to that
user. Authors propose in \cite{influence-cookbooks} a graph theoretic
approach to retrieve which \textit{centrality} metrics are being used
by peer-ranking platforms. Work in \cite{stealing} shows that machine
learning models can be extracted by a user from the cloud platform
where they are executed, in order to leverage the leaked model to issue
new predictions. Reference \cite{10.1145/3308558.3313674} observes
online auction systems.  In the domain of recommender systems, paper
\cite{Sinha01beyondalgorithms:} exposes the users perspective on what
they expect from recommendation.

\paragraph{Observation and statistics in Twitter}
The specific case of Twitter was consistently studied for multiple
research leads, including in the INFOCOM community \cite{ego,6848048}.
Recently, Gilani et al. \cite{10.1145/3298789} study the behavioral
characteristics of both populations of bots and humans in Twitter.
Twitter data is often represented as graphs in order to extract
relevant information, such as relationship structures \cite{ego}, the
``follower'' mechanism \cite{10.1145/2567948.2576939}, general
dynamics \cite{6848048}, or influencers \cite{ZENGINALP2018211}.
 
\vspace{0.2cm}
\section{Conclusion}
\label{sec:conclusion}

Allegations of shadow banning practices have been countless in the
media and the population in the recent years. Yet, no objective
approach ever quantified this practice. We proposed in this paper to
remedy this lack, by observing at large scale shadow banning practices
on a major online social network. We then presented statistical
approaches leveraging the collected dataset to shed light on the
phenomenon.

First, we explored public Twitter-user features to seek a relation
between these features and ban statuses.  Then, through two
statistical modeling hypotheses, we compared the likelihood of two
narratives commonly encountered around shadow ban questions. Our
conclusions indicate that bans appear as a local event, impacting
specific users and their close interaction partners, rather than
resembling a (uniform) random event such as a bug.

As of future work, we believe one crucial notion to be analyzed is the
temporal dimension of the shadow banning phenomenon: \eg how does a
shadow ban status evolves among neighbors? Can the beginning of a ban
be correlated in time with other observables? Is the appearance of
the shadow ban statuses mostly happening in batch, or is the
propagation smooth among the monitored user profiles? Another
important observation to be conducted is the possible reversibility of
this status: can we observe user profiles retrieving their initial
visibility (\ie losing the shadow ban status they had), after they for
instance interacted less with shadow banned users?  Lastly, we have
chosen to use both a statistical and topological approach in our
study; there are probably several other interesting approaches to
address shadow banning under other angles, for instance at the
semantic level by analyzing the contents of the messages. We think
these other interesting dimensions to be of great interest for
scientists, algorithm designers and the general public.

\section{Data and Code Availability Statement}
\label{sec:data}

We release an anonymized version of the dataset we gathered for this study, as well as the
code for our core experiments, at the following location: \url{https://gitlab.enseeiht.fr/bmorgan/infocom-2021}.

\section{Acknowledgements}
We thank the \textit{shadowban.eu} initiative, providing tests and code to
individuals for spotting shadow banning practices.

  \bibliographystyle{IEEEtran}
  \bibliography{biblio}

\end{document}